# A Taxonomy of Data Risks in AI and Quantum Computing (QAI): A Systematic Review

Grace Billiris
School of Computer Science
University of Technology, Sydney
Australia
grace.v.billiris@student.uts.edu.au

Asif Gill
School of Computer Science
University of Technology, Sydney
Australia
asif.gill@uts.edu.au

Madhushi Bandara
School of Computer Science
University of Technology, Sydney
Australia
madhushi.bandara@uts.edu.au

## Abstract

*Quantum Artificial Intelligence (QAI), integrating Artificial Intelligence (AI) and Quantum Computing (QC), promises transformative advances, such as enhancing privacy though AI-enabled quantum cryptography or securing future digital systems with quantum-resistance encryption protocols. A key barrier for QAI applications is that they inherit data risks associated with both AI and QC, including privacy and security concerns. AI's reliance on sensitive datasets, combined with QC's threat to classical encryption, creates complex vulnerabilities that are not studied systematically. These issues directly impact the trustworthiness and reliability of AI or QAI systems, making them critical to address. This study aims to expand understanding by systematically reviewing 67 privacy and security-related studies. Our main contribution is a taxonomy developed to classify QAI data risks systematically. We identify 22 key data risks grouped into five categories: governance, risk assessment, control implementation, user considerations, and continuous monitoring. Our findings reveal emerging vulnerabilities unique to QAI and highlight significant gaps in holistic understanding. This paper contributes to advancing trustworthy AI or QAI and provides a future research and risk assessment tool development to manage the evolving QAI data risk landscape.*

## 1. Introduction

Quantum Artificial Intelligence (QAI) is an emerging paradigm that combines the capabilities of Artificial Intelligence (AI) and Quantum Computing (QC) (Huang et al., 2022; Hadap et al., 2024). AI systems enable machines to perform tasks such as learning and reasoning using massive datasets and computing resources (e.g., CPUs, GPUs) (Billiris et al., 2024; Lee et al., 2024; Wirtz, 2024; Ali et al., 2024; Madanchian & Taherdoost, 2024; Majeed et al., 2022; Hadap et al., 2024). QC, on the other hand, harnesses quantum mechanics to solve problems at speeds unattainable by classical computing (ISACA, 2022; Alaeifar et al., 2024). Both AI and QC individually have the potential to revolutionise multiple sectors such as healthcare, defence, finance, and cybersecurity (Muheidat et al., 2024; National Institute of Standards and Technology, 2023; Manjunath & Bhowmik, 2023; Majeed et al., 2022; Theofanos et al., 2024; Vasani et al., 2024; Madanchian & Taherdoost, 2024; Konstantinova et al., 2022; Autio et al., 2024). Similarly, QAI is poised to drive transformative change across these domains by integrating the strengths of both AI and QC (Lakhyar, 2024). However, the combination of these technologies also inherits the data risks associated with each, including both data privacy and data security concerns (Lakhyar, 2024; Hadap et al., 2024; Feretzakis et al., 2024). These issues directly impact the trustworthiness and reliability of AI or QAI systems, making them critical to address.

In the context of our study, data privacy refers to protecting individuals' rights over their personal data, focusing on consent, transparency, and user control (Majeed et al., 2022; Wakili & Bakkali, 2025). Data security involves the technical measures that safeguard data from unauthorised access or compromise, such as encryption and access controls. AI systems are increasingly processing sensitive personal data (Verma, 2024; Ye et al., 2024; CEDPO AI Working Group, 2023; Vardalachakis et al., 2024; Ali et al., 2024; Saeed & Alsharidah, 2024). At the same time, quantum algorithms threaten to undermine traditional encryption methods such as Rivest-Shamir-Adleman (RSA) and Advanced Encryption Standard (AES) (Ali et al., 2024; Wakili & Bakkali, 2025). For example, Shor's algorithm demonstrates the feasibility of breaking these cryptographic standards (Vasani et al., 2024; Manjunath & Bhowmik, 2023; Baseri et al., 2024; Ali et al., 2024), creating an urgent need for quantum-resilient privacy and security techniques (Ali et al., 2024; Wadhwa, 2022; LaMacchia, 2022; Feretzakis et al., 2024). These challenges highlight the need to synthesise data risks – encompassing both privacy and security – in AI and QC, that can lay a foundation for understanding and addressing the data risk landscape for QAI (Hiwale et al., 2023; Feretzakis et al., 2024; Lee et al., 2024; Li et al., 2020; Lakhyar, 2024; Spoorthi, 2024). Addressing these risks contributes to advancing trustworthy AI or QAI, ensuring that future QAI systems are reliable and secure.

This study addresses this need through a systematic literature review, guided by the research question: What are the key data risks in QAI?

We followed the PRISMA 2020 method (PRISMA, 2020) to ensure a transparent and rigorous selection process. This review offers a structured synthesis of dispersed literature, casting the findings into a comprehensive and actionable taxonomy that serves as a knowledge base for both researchers and practitioners. Our analysis also identifies critical gaps in current privacy and security research and highlights underexplored risks emerging from QAI. These insights provide a solid foundation for future researchers in design and development of QAI data risk assessment tool and mitigation strategies.

This paper is organised as follows: Section 2 outlines the research background and motivation, followed by the methodology in Section 3. Sections 4 to 6 present our findings, discussion, and conclusions with future directions respectively.

## 2. Research Background and Motivation

### 2.1. Artificial Intelligence and Data Privacy Risks

AI refers to the simulation of human intelligence in machines that perform cognitive tasks such as learning, reasoning, and decision-making (Billiris et al., 2024; Lee et al., 2024; Wirtz, 2024; Ali et al., 2024; Madanchian & Taherdoost, 2024; Majeed et al., 2022; Hadap et al., 2024). While AI encompasses a broad range of techniques, including rule-based systems and symbolic reasoning, many contemporary AI applications—particularly those based on machine learning, deep learning, and other statistical models—rely heavily on vast datasets to perform functions such as prediction, classification, and decision-making (Alkaeed et al., 2024; Feretzakis et al., 2024; Spoorthi, 2024; Wirtz, 2024; Konstantinova et al., 2022; Hadap et al., 2024; Vardalachakis et al., 2024; Verma, 2024; Majeed et al., 2022; Vassilev et al., 2024). These data dependency raises significant privacy concerns, as these datasets often contain sensitive personal or organisational information, including health records, financial data, or behavioural patterns (Ye et al., 2024; Vardalachakis et al., 2024; Verma, 2024; Ali et al., 2024; Feretzakis et al., 2024; Saeed & Alsharidah, 2024). Additionally, the lack of model transparency in many AI systems makes it difficult to assess how data is processed, or decisions are made, further complicating efforts to ensure accountability and user privacy (Saeed & Alsharidah, 2024). While the performance of AI systems improves with increased data access (CEDPO AI Working Group, 2023), this reliance also significantly heightens the risk of data breaches (Lee et al., 2024; Feretzakis et al., 2024; CEDPO AI Working Group, 2023; Saeed & Alsharidah, 2024; Verma, 2024; Vardalachakis et al., 2024; Saeed & Alsharidah, 2024).

### 2.2. Quantum Computing and Data Privacy Risks

QC introduces a new computational paradigm that promises to exponentially accelerate problem-solving capabilities (Choppakatla, 2023; Cheah, 2024; Byeon et al., 2025; Ramu et al., 2022; Abdelgaber & Nikolopoulos, 2020; Gill & Buyya, 2024; Alexeev et al., 2024). While this holds potential to deliver better AI system performance and accelerated scientific discovery, it also poses a direct threat to traditional data cryptographic methods (Vasani et al., 2024; Gill & Buyya, 2024; Guo, 2021). Quantum algorithms like Shor's and Grover's can compromise widely used encryption schemes such as RSA and AES, rendering them obsolete in a post-quantum world (Vasani et al., 2024; Manjunath & Bhowmik, 2023; Baseri et al., 2024; Ali et al., 2024; Saeed & Alsharidah, 2024; Wakili & Bakkali, 2025).

This advancement necessitates the development of Post-Quantum Cryptography (PQC) and Quantum Key Distribution (QKD), and existing technologies in this domain remain immature, with limited standardisation and adoption (Baseri et al., 2024; Wadhwa, 2022; LaMacchia, 2022; Feretzakis et al., 2024; Ali et al., 2024).

### 2.3. QAI and Data Privacy Risks

QAI introduces a layered set of privacy challenges (Hadap et al., 2024; Lakhyar, 2024). QC can amplify risks pose by AI such as data dependency and model transparency issues through its cryptographic-breaking capabilities (Lakhyar, 2024; Ali et al., 2024; Hadap et al., 2024; Saeed & Alsharidah, 2024). Together, these technologies not only create new vulnerabilities–such as cryptographic comprise though quantum algorithms and large-scale data leakage from increased data reliance–but also exacerbate existing like insufficient encryption safeguards and lack of explainability in AI systems (Hiwale et al., 2023; Feretzakis et al., 2024). Their integration into existing systems is complex, especially in environments where AI models already struggle with computational constraints. (Lakhyar, 2024; Alexeev et al., 2024; Huang et al., 2022; Vasani et al., 2024; Mangla et al., 2023; Kwak et al., 2022; Wang & Tang, 2024; Singh et al., 2024; Satuluri & Ponnusamy, 2021; Alexeev et al., 2024)

Despite increasing interest in QAI, the literature exploring its data risks remains fragmented (Zhang et al., 2024; Lakhyar, 2024). Most existing studies and regulations focus on either AI or QC in isolation, ignoring the risk profile that emerges from their combination (ISACA, 2025; Lakhyar, 2024; Golda et al., 2024; Spoorthi, 2024). This gap in the literature underscores the need for a systematic literature review and development of a data risk taxonomy that systematically organises the diverse data risks arising from QAI.

# 3. Research Method

This study investigates the research question: What are the key data risks in QAI?

To answer this question, we conducted a systematic literature review using the PRISMA (PRISMA, 2020) methodology to identify, synthesise, and classify data risks and mitigation strategies in QAI, focusing on literature published between 2020 and 2025.

Searches were performed across seven databases: IEEE Xplore (https://ieeexplore.ieee.org/), ACM Digital Library (https://www.acm.org/), ScienceDirect (https://www.sciencedirect.com/), ResearchGate (https://www.researchgate.net/), National Institute of Standards and Technology (NIST) (https://www.nist.gov/), Gartner (https://www.gartner.com/), and Information Systems Audit and Control Association (ISACA) (https://www.isaca.org/).

The Boolean search string applied was: *"data risks" AND (("AI" OR "Artificial Intelligence") OR ("QC" OR "Quantum Solutions" OR "Quantum Software" OR "Quantum System" OR "Quantum Computing")).*

## 3.1. Study Selection and Quality Assessment

A five-stage selection process was followed as illustrated in Figure 1, beginning with the initial identification (1,388 studies). Duplicate studies were removed before title/keyword screening, reducing the pool to 399 studies, followed by the abstract/conclusion screening (157 studies), and full-text review (100 studies). Snowballing techniques were integrated throughout these stages to identify additional relevant studies. The process concluded with 67 studies included in the final analysis.

To ensure transparency and methodological rigour, predefined inclusion and exclusion criteria were applied as follows:

Inclusion criteria:

- Studies addressing data risks in AI, QC, or their intersection.
- Peer-reviewed articles published in English between 2020 and 2025.

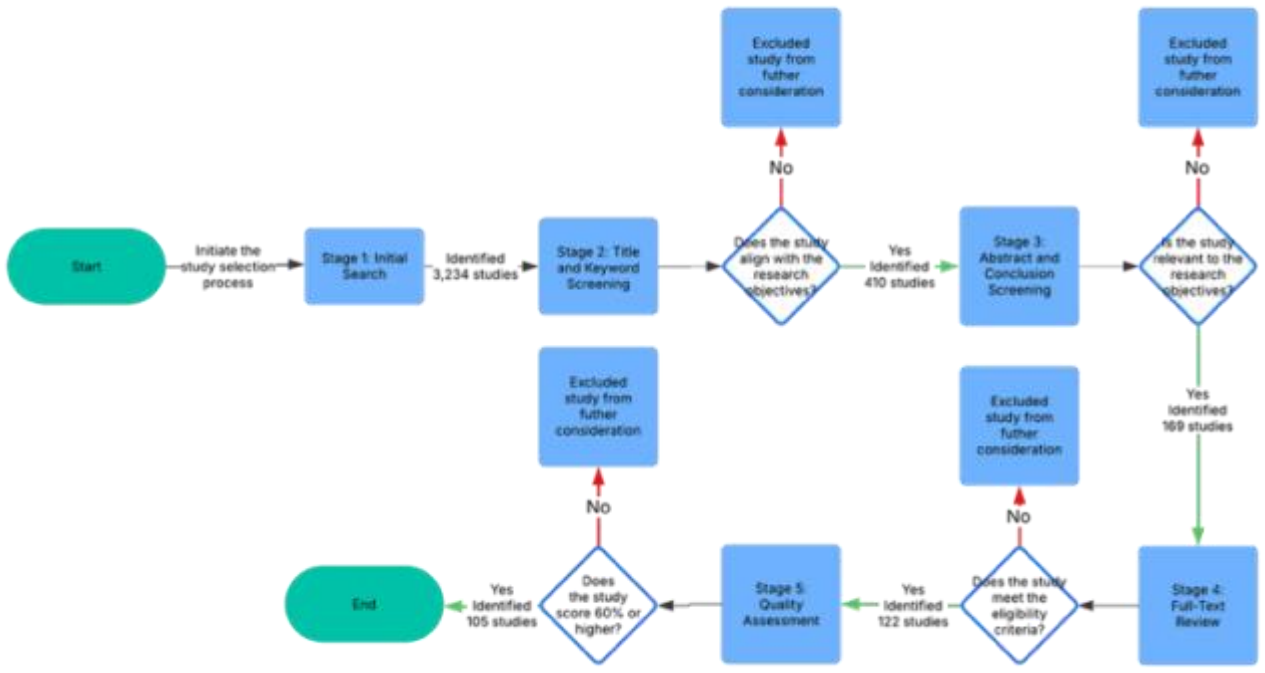

**Figure 1: Flow Diagram of the Selection Process Based on PRISMA (2020).**

- Peer-reviewed articles, conference papers, technical reports, white papers, and institutional publications (e.g., NIST, ISACA).
- Studies exploring privacy-preserving techniques, quantum solutions, artificial intelligence, or frameworks addressing privacy risks.

Exclusion criteria:

- Studies unrelated to AI or QC privacy concerns.
- Non-peer-reviewed publications or inaccessible full texts.
- Research lacking methodological clarity or detailed findings.

Quality assessment used a checklist adapted from (Kitchenham & Charters, 2007) to evaluate methodological rigour, clarity, and relevance. Studies were scored on criteria such as clarity of objectives, transparency of methods, validity of findings, and relevance to AI/QC privacy intersections. Only studies scoring 3/5 or above were included in the final synthesis, balancing the inclusion of robust research while allowing valuable exploratory or theoretical contributions. Appendix A contains full list of identified studies including the results of the quality assessment.

### 3.2 Taxonomy Development and Validation

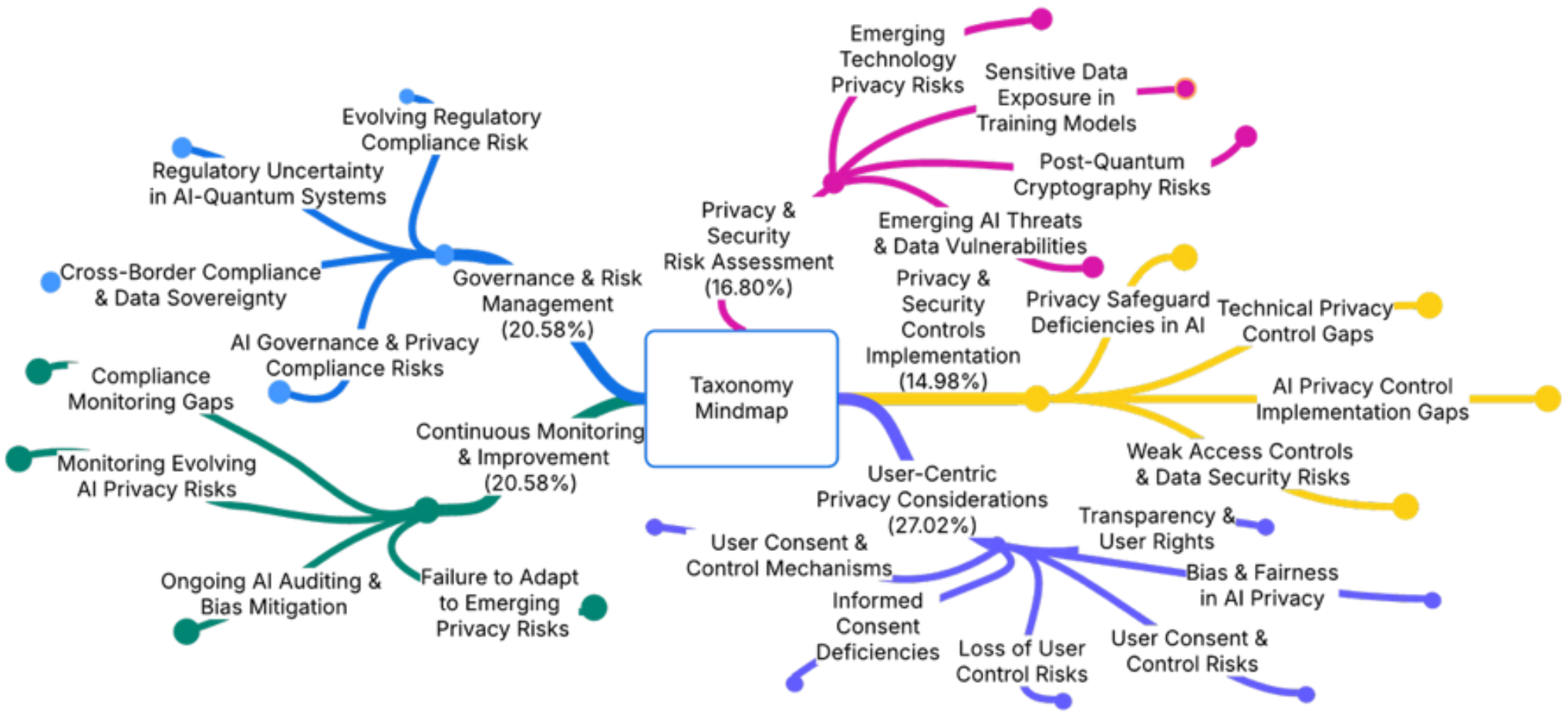

**Figure 2: Taxonomy of Data Privacy Risks.**

To systematically categorise the data risks identified through our review, we developed a five-branch taxonomy grounded in thematic analysis. The process involved several key stages:

1. Thematic Coding of Risks

Following full-text review, we performed inductive coding across the 67 included studies to extract specific data privacy and security risks. These risks were grouped based on recurring patterns in terminology, concepts, and focus areas. This stage produced 22 distinct risk types.

2. Iterative Categorisation into Thematic Branches

The identified risks were then clustered into five overarching thematic categories; User-Centric Privacy Considerations, Privacy & Security Risk Assessment, Privacy & Security Controls Implementation, Governance & Risk Management and Continuous Monitoring & Improvement. These themes were informed by both domain knowledge and common risk typologies across AI, QC, privacy and security literature. Colour-coded branches were applied to the taxonomy to visually distinguish each thematic grouping (as depicted in Figure 2).

3. Validation via Review

To support clarity and consistency, the taxonomy underwent fortnightly review sessions with senior researchers. Each risk definition and thematic grouping was revisited to assess its relevance to QAI and alignment with the inclusion criteria. Where needed, adjustments were made through discussion and consensus to maintain coherence across categories and terminology.

4. Consistency and Terminology

The taxonomy was refined to use consistent terminology drawn from the literature reviewed. Risk definitions and category labels were adjusted for clarity and to reflect common language used in related privacy and security work.

This structured process allowed us to produce a well-defined taxonomy of QAI data risks, which underpins the analysis presented in Section 4 and detailed in Appendix B.

## 4. Results and Findings

Using the 67 studies we selected, the thematic analysis revealed 22 key risks, and five thematic categories of them. The mapping of identified risks into one of five thematic categories is represented in Figure 2. This categorisation and mapping structures the identified risks in a systematic manner, facilitating a clearer understanding of their nature and impact. A detailed summary of these risks can be found in Appendix B.

Of the 22 identified data risks, 18 (82%) relate primarily to data privacy concerns—such as user consent and control, transparency and user rights—while 4 (18%) pertain to data security aspects including weak access controls and data security risks (see Appendix B for all privacy/security risks). This distribution reflects a higher frequency of privacy-related risks compared to security risks in the QAI context.

We further analysed the technological scope of the 67 studies reviewed:

- 34 studies (50.7%) focused on data risks related to AI systems.

- 29 studies (43.3%) examined risks in QC contexts.
- Only 4 studies (6%) addressed QAI specifically.

This distribution indicates that while AI and QC are individually well-represented in current literature, QAI as a combined field remains relatively underexplored, despite its compounded risk landscape.

The following sections provide key findings under the five thematic categories. A comprehensive list of risks identified under each category is detailed in Appendix B.

## 4.1. Governance and Risk Management

20.58% of the reviewed studies highlight critical weaknesses in data governance, regulatory compliance, and organisational alignment with evolving privacy standards, revealing a pervasive inability to keep pace with emerging technological demands. For instance, S44 and S22 specifically point to fragmented governance structures that undermine consistent policy enforcement, while S36 and S14 emphasise gaps in regulatory frameworks that leave organisations ill-prepared to address novel risks introduced by AI and quantum technologies. The most frequently cited risk, noted by S36, S14, and I4, concerns the continued reliance on inconsistent or outdated privacy policies, which fail to incorporate the nuanced challenges of emerging technologies, thereby increasing organisational exposure to compliance breaches and privacy violations. This collective evidence underscores a significant misalignment between existing governance mechanisms and the rapidly evolving technological landscape.

## 4.2. Privacy & Security Risk Assessment

Accounting for 16.8% of all identified risks, studies such as S4, S35, and S32 critically highlight vulnerabilities in organisational approaches to evaluating and managing data risks within QAI systems, revealing gaps in both risk identification and mitigation processes. Notably, the exposure of personal data during AI model training and inference emerges as a predominant concern, as underscored by I1, S10, and S26, who demonstrate how inadequate data handling and insufficient anonymisation techniques can lead to significant privacy breaches. Further, S46, and S55 emphasise that these vulnerabilities are often exacerbated by the complexity and opacity of emerging technologies such as AI, QC, or their convergence in QAI, which complicate effective data risk assessment and control.

## 4.3. Privacy & Security Controls Implementation

14.98% of the reviewed studies (e.g. S29, S34, S52, and S23) underscore critical deficiencies in the technical safeguards implemented across AI, QC, and QAI systems—particularly in high-stakes sectors such as healthcare and finance, where data sensitivity is heightened. A recurring and high-priority concern is the prevalence of weak or misconfigured access controls, as highlighted by S18, S37, S47, and S49. These studies demonstrate how inadequate enforcement of access policies not only exposes systems to unauthorised data but also undermines broader security postures in environments already challenged by rapid technological change.

## 4.4. User-Centric Privacy Considerations

27.02% of the reviewed studies identify a persistent erosion of user trust and a lack of meaningful user agency in the design and deployment of AI, QC, and QAI systems. Several studies—including S46, S42, S40, and S28—critique the limited opportunities for users to understand, influence, or challenge how their data is collected and used. A particularly salient risk, noted in S39, S26, and S10, is the absence of robust informed consent mechanisms, which undermines transparency and user control in AI-driven data processing. These limitations not only jeopardise individual rights but also weaken the social legitimacy of emerging technologies.

## 4.5. Continuous Monitoring and Improvement

20.58% of the reviewed studies expose substantial deficiencies in the continuous evaluation and adaptation of data protection practices across AI, QC, and QAI systems, as noted in studies such as S4, S9, and S19. These studies highlight a reactive rather than proactive approach to data governance, where static policies fail to keep pace with dynamic risk landscapes. The most frequently cited shortcoming—emphasised by S15 and I4—is the absence of real-time data auditing mechanisms, which hinders the timely detection and mitigation of emerging threats.

## 4.6 Summary

Our taxonomy and findings (Figure 2 and Appendix B) reflect a multidimensional landscape of data risks emerging across AI, QC, and QAI systems. The reviewed studies reveal systemic gaps in governance, technical safeguards, user-centric protections, and ongoing oversight—underscoring the complexity of managing privacy and security in these evolving domains. Many of these risks parallel those identified in broader initiatives, such as the MIT AI Risk Project (Massachusetts Institute of Technology, 2025), reinforcing their relevance and urgency. In the following section, we examine how these insights deepen theoretical understanding and inform ongoing efforts to improve the trustworthiness and reliability of AI or QAI systems and develop practical strategies for managing emerging data risks in QAI contexts.

# 5. Discussion

The findings of this systematic review highlight the urgent need for tailored data privacy and security approaches specific to emerging technologies such as AI, QC, and their convergence in QAI, which are essential for building trustworthy and reliable AI or QAI systems. The developed taxonomy provides a structured approach to classifying and interpreting the unique data risks inherent in these domains. In doing so, it directly addresses the research question by elucidating how such risks manifest in QAI environments and identifying critical gaps in current approaches to risk understanding and management.

## 5.1. Key Insights

The taxonomy derived from this review offers a classification of data risks that is not only comprehensive but actionable. By mapping each risk to governance structures, technical controls, user expectations, and continuous improvement mechanisms, the taxonomy intends to assist practitioners and policymakers to identify, analyse, and categorise data risks and mitigation strategies from different stakeholder perspectives. Key insights include:

- Governance challenges are widespread, with fragmented structures and outdated policies failing to keep pace with emerging AI and quantum technologies, increasing exposure to compliance and privacy breaches.
- Risk assessment processes often lack depth and fail to adequately identify vulnerabilities in QAI systems, especially concerning data exposure during AI model training and inference.
- Technical controls show critical weaknesses, particularly in access control enforcement, with many systems vulnerable due to misconfiguration or insufficient safeguards.
- User-centric concerns—such as consent, fairness, and transparency—remain under prioritised in technical implementations.

These findings confirm that current risk management approaches, whether technical or organisational, are not yet equipped to handle the layered and evolving risks posed by QAI systems.

## 5.2. Implications

This study provides both theoretical and practical implications. Theoretically, it provides a structured research-based model to classify data risks in QAI. These findings can inform the design of privacy and security-enhancing artifacts that address the challenges of QAI. Additionally, the taxonomy indicates regulatory, technical, and ethical considerations in QAI, offering a reusable classification model for researchers exploring hybrid data risk domains.

In practice, practitioners can apply the taxonomy to develop tailored data risk assessments for QAI systems. The taxonomy supports the evaluation of an organisation's internal privacy posture against QAI-specific risk categories and facilitates gap analyses to ensure alignment with evolving data protection regulations.

## 5.3. Limitations and Future Work

This review acknowledges several limitations. The search strategy was based on peer-reviewed sources across databases including IEEE Xplore, ACM, NIST, ResearchGate, ScienceDirect, Gartner and ISACA. While this ensured academic and industry relevance, it may have excluded emerging grey literature or less formal insights.

The final selection included 67 high-quality studies, refined through a rigorous inclusion and quality assessment process. However, the interpretation of data risks is inherently subjective, and classification decisions—although based on existing literature—carry some bias.

Lastly, this review provides a theoretical lens; empirical validation remains essential. Future research should apply taxonomy in real-world QAI systems to assess its effectiveness. There is also a need for the development and evaluation of a risk assessment tool specifically designed to mitigate data privacy and security risks arising from QAI.

# 6. Conclusion

This study presents a systematic review of data risks emerging from AI, QC, and QAI. By analysing 67 high-quality studies, we identified 22 key data risks that span emergent themes: governance, risk assessment, control implementation, user considerations and continuous monitoring. These risks reflect both the amplification of existing challenges and the emergence of novel threats introduced by the convergence of AI and quantum technologies. These issues directly impact the trustworthiness and reliability of AI or QAI systems, making them critical to address.

The taxonomy developed through this review contributes to academic discourse by offering a structured classification lens that integrates regulatory, technical, and human-centric considerations. It also provides a practical foundation for risk identification, assessment, and strategic planning tailored to QAI contexts. Ultimately, this work contributes to advancing trustworthy AI by equipping researchers and practitioners with the necessary tools to understand and mitigate emerging data risks in QAI systems.

## 7. Acknowledgements

This research is supported by an Australian Government Research Training Program Scholarship. The authors gratefully acknowledge this support, which has aided the completion of this paper.

# 9. Appendix

## A – Selected Studies Based on Quality Assessment

Note: I = Industry, S = (Academic) Study.

| *Study Title* | *Study Num.* | *Research* | *Aim* | *Context* | *Finding* | *Future* | *Total Score* |
|---|---|---|---|---|---|---|---|
| *Securing the Future: Mitigating Data Security Concerns in AI Models* | I1 | 1 | 1 | 0 | 1 | 0 | 3 |
| *Demystifying Quantum* | I2 | 1 | 1 | 1 | 1 | 0 | 4 |
| *The Quantum Computing Threat: Risks and Responses* | I3 | 1 | 1 | 1 | 1 | 1 | 5 |
| *State of Privacy 2025* | I4 | 1 | 1 | 1 | 1 | 1 | 5 |
| *How to Mitigate the Cryptography Risks Posed by Quantum Computing* | I5 | 1 | 1 | 1 | 1 | 1 | 5 |
| *Predicts 2025: Privacy in the Age of AI and the Dawn of Quantum* | I6 | 1 | 1 | 1 | 1 | 1 | 5 |
| *Quantum Machine Learning: Bridging the Gap Between Quantum Computing and Artificial Intelligence: An Overview* | S1 | 1 | 1 | 1 | 1 | 1 | 5 |
| *Artificial Intelligence for Quantum Error Correction: A Comprehensive Review* | S2 | 1 | 1 | 1 | 1 | 1 | 5 |
| *Artificial Intelligence and Quantum Computing* | S3 | 1 | 1 | 1 | 1 | 1 | 5 |
| *Quantum Collaboration: Pioneering the Frontier of Enhanced Artificial Intelligence* | S4 | 1 | 1 | 1 | 1 | 1 | 5 |
| *A Revolution in Processing Capabilities and Its Possible Uses: Quantum Computing* | S5 | 1 | 1 | 1 | 1 | 1 | 5 |
| *Privacy-preserving security of IoT networks: A comparative analysis of methods and applications* | S6 | 1 | 1 | 1 | 1 | 1 | 5 |
| *Overview on Quantum Computing and its Applications in Artificial Intelligence* | S7 | 1 | 1 | 1 | 1 | 1 | 5 |
| *Current approaches and future directions for Cyber Threat Intelligence sharing: A survey* | S8 | 1 | 1 | 1 | 1 | 1 | 5 |
| *When AI Meets Information Privacy: The Adversarial Role of AI in Data Sharing Scenario* | S9 | 1 | 1 | 1 | 1 | 1 | 5 |
| *Generative AI: The Data Protection Implications* | S10 | 1 | 1 | 1 | 1 | 1 | 5 |
| *Quantum Machine Learning and Recent Advancements* | S11 | 1 | 1 | 1 | 1 | 1 | 5 |
| *An Introduction To Quantum Machine Learning Techniques* | S12 | 1 | 1 | 1 | 1 | 1 | 5 |
| *Analysis on the recent development of quantum computer and quantum neural network technology* | S13 | 1 | 1 | 1 | 1 | 1 | 5 |
| *Privacy-Preserving Techniques in Generative AI and Large Language Models: A Narrative Review* | S14 | 1 | 1 | 1 | 1 | 1 | 5 |
| *Generative AI and Data Privacy Concerns* | S15 | 1 | 1 | 1 | 1 | 1 | 5 |
| *Research on Differential Privacy Protection of Power Grid Data Based on Artificial Intelligence and Federated Learning* | S16 | 1 | 1 | 1 | 1 | 1 | 5 |
| *The MIT AI Risk Repository* | S17 | 1 | 1 | 1 | 1 | 1 | 5 |
| *Federated Quantum-Based Privacy-Preserving Threat Detection Model for Consumer Internet of Things* | S18 | 1 | 1 | 1 | 1 | 1 | 5 |
| *Why Quantum Computing Is Even More Dangerous Than Artificial Intelligence* | S19 | 1 | 1 | 1 | 1 | 1 | 5 |
| *Quantum Computing: Threats & Possible Remedies* | S20 | 1 | 1 | 1 | 1 | 1 | 5 |
| *Machine learning enabling high-throughput and remote operations at large-scale user facilities* | S21 | 1 | 1 | 1 | 1 | 1 | 5 |
| *Empirical Evaluation of Federated Learning with Local Privacy for Real-World Application* | S22 | 1 | 1 | 1 | 1 | 1 | 5 |
| *Privacy and Security Concerns in Generative AI: A Comprehensive Survey* | S23 | 1 | 1 | 1 | 1 | 1 | 5 |
| *Toward Privacy Preservation Using Clustering Based Anonymization: Recent Advances and Future Research Outlook* | S24 | 1 | 1 | 1 | 1 | 1 | 5 |
| *Quantum-Enhanced Machine Learning* | S25 | 1 | 1 | 1 | 1 | 1 | 5 |
| *The Future of Privacy: A Review on AI's Role in Shaping Data Security* | S26 | 1 | 1 | 1 | 1 | 1 | 5 |
| *Panel Summary Report: Risk Management & Industrial Artificial Intelligence, INFORMS Annual Meeting, 2021 (Virtual)* | S27 | 1 | 1 | 1 | 1 | 1 | 5 |
| *Explainable AI for cybersecurity automation, intelligence and trustworthiness in digital twin: Methods, taxonomy, challenges and prospects* | S28 | 1 | 1 | 1 | 1 | 1 | 5 |
| *Federated Learning for Data Security and Privacy Protection* | S29 | 1 | 1 | 1 | 1 | 1 | 5 |
| *Towards the Development of a Copyright Risk Checker Tool for Generative Artificial Intelligence Systems* | S30 | 1 | 1 | 1 | 1 | 1 | 5 |
| *Navigating Quantum Security Risks in Networked Environments: A Comprehensive Study of Quantum-Safe Network Protocols* | S31 | 1 | 1 | 1 | 1 | 1 | 5 |
| *The long road ahead to transition to post-quantum cryptography* | S32 | 1 | 1 | 1 | 1 | 1 | 5 |
| *Quantum Computing in Artificial Intelligence: A Paradigm Shift* | S33 | 1 | 1 | 1 | 1 | 1 | 5 |
| *Adversarial Machine Learning: A Taxonomy and Terminology of Attacks and Mitigations* | S34 | 1 | 1 | 1 | 1 | 1 | 5 |
| *Challenge Design and Lessons Learned from the 2018 Differential Privacy Challenges* | S35 | 1 | 1 | 1 | 1 | 1 | 5 |
| *Deepfakes, Phrenology, Surveillance, and More! A Taxonomy of AI Privacy Risks* | S36 | 1 | 1 | 1 | 1 | 0 | 4 |
| *Security of federated learning in 6G era: A review on conceptual techniques and software platforms used for research and analysis* | S37 | 1 | 1 | 1 | 1 | 1 | 5 |
| *Blockchain and AI technology convergence: Applications in transportation systems* | S38 | 1 | 1 | 1 | 1 | 1 | 5 |
| *AI Risk Management Framework* | S39 | 1 | 1 | 1 | 1 | 1 | 5 |
| *Security, privacy, and robustness for trustworthy AI systems: A review* | S40 | 1 | 1 | 1 | 1 | 1 | 5 |
| *Privacy-preserving explainable AI enable federated learning-based denoising fingerprint recognition model* | S41 | 1 | 1 | 1 | 1 | 1 | 5 |
| *Artificial Intelligence Risk Management Framework: Generative Artificial Intelligence Profile* | S42 | 1 | 1 | 1 | 1 | 1 | 5 |
| *QLSN: Quantum key distribution for large scale networks* | S43 | 1 | 1 | 1 | 1 | 1 | 5 |
| *A systematic review of privacy-preserving methods deployed with blockchain and federated learning for the telemedicine* | S44 | 1 | 1 | 1 | 1 | 1 | 5 |
| *Quantum computation in power systems: An overview of recent advances* | S45 | 1 | 1 | 1 | 1 | 1 | 5 |
| *AI-driven fusion with cybersecurity: Exploring current trends, advanced techniques, future directions, and policy implications for evolving paradigms– A comprehensive review* | S46 | 1 | 1 | 1 | 1 | 1 | 5 |
| *Privacy preservation in Artificial Intelligence and Extended Reality (AI-XR) metaverses: A survey* | S47 | 1 | 1 | 1 | 1 | 1 | 5 |
| *Quantum distributed deep learning architectures: Models, discussions, and applications* | S48 | 1 | 1 | 1 | 1 | 1 | 5 |
| *Security and privacy of internet of medical things: A contemporary review in the age of surveillance, botnets, and adversarial ML* | S49 | 1 | 1 | 1 | 1 | 1 | 5 |
| *Privacy preservation in Distributed Deep Learning: A survey on Distributed Deep Learning, privacy preservation techniques used and interesting research directions* | S50 | 1 | 1 | 1 | 1 | 1 | 5 |
| *A Foundation for Investigating Practitioner Perspective"* | S51 | 1 | 1 | 1 | 1 | 0 | 4 |
| *Embracing the quantum frontier: Investigating quantum communication, cryptography, applications and future directions* | S52 | 1 | 1 | 1 | 1 | 1 | 5 |
| *Federated learning enabled digital twins for smart cities: Concepts, recent advances, and future directions* | S53 | 1 | 1 | 1 | 1 | 1 | 5 |
| *Quantum-centric supercomputing for materials science: A perspective on challenges and future directions* | S54 | 1 | 1 | 1 | 1 | 1 | 5 |
| *Privacy and Personal Data Risk Governance for Generative Artificial Intelligence: A Chinese Perspective* | S55 | 1 | 1 | 1 | 1 | 1 | 5 |
| *Artificial Intelligence for Quantum Computing* | S56 | 1 | 1 | 1 | 1 | 1 | 5 |
| *AI Use Taxonomy: A Human-Centered Approach* | S57 | 1 | 1 | 1 | 1 | 1 | 5 |
| *Transforming Research with Quantum Computing* | S58 | 1 | 1 | 1 | 1 | 1 | 5 |
| *NIST Privacy Framework: A Tool for Improving Privacy Through Enterprise Risk Management, Version 1.0* | S59 | 1 | 1 | 1 | 1 | 1 | 5 |
| *AI-Powered Innovations in High-Tech Research and Development: From Theory to Practice* | S60 | 1 | 1 | 1 | 1 | 1 | 5 |
| *Applying AI and Machine Learning to Enhance Automated Cybersecurity and Network Threat Identification* | S61 | 1 | 1 | 1 | 1 | 1 | 5 |

## B – Overview of Data Privacy Risks in AI and Quantum Computing

To support interpretation of the "Frequency" column in the table below; frequency refers to how often a particular risk appears across the reviewed studies, expressed using both count and percentage values:

- Count: The number of studies in which the risk was identified.
- Percentage: The proportion of this count relative to the total number of risks mentions across all categories.

Note: P = Privacy, S = Security.

| Identified Risk | Studies | Frequency | Definition | P/S Category |
|---|---|---|---|---|
| | | **Category: Governance & Risk Management** | | |
| **AI Governance & Privacy Compliance Risks** | I1–I6, S1–S61 | Count: 50; Percentage: 7% | Risks associated with the alignment of AI governance frameworks with privacy compliance standards, leading to inadequate privacy protections. | P |
| **Evolving Regulatory Compliance Risks** | I3–I6, S1–S5, S7, S10, S12–S15, S17–S19, S25, S27–S32, S33–S34, S38–S40, S42, S43, S45–S46, S49, S52–S56, S58 | Count: 39; Percentage: 5.46% | Risks arising from changing regulatory requirements that may leave an organisation non-compliant with new laws or privacy standards. | P |
| **Cross-Border Compliance & Data Sovereignty** | I3, I5–I6, S2–S10, S12–S16, S17–S23, S26–S33, S34–S40, S41–S61 | Count: 30; Percentage: 4.2% | Risks related to non-compliance with varying data protection laws across different countries, impacting international data flow and sovereignty. | P |
| **Regulatory Uncertainty in AI-Quantum Systems** | I2, I3, I5–I6, S1–S7, S11–S13, S20–S23, S25–S34, S37–S41, S43, S44, S48, S51, S54–S59 | Count: 28; Percentage: 3.92% | Risks from unclear or inconsistent regulations surrounding the integration of AI with quantum systems, potentially leading to privacy lapses. | P |
| | | **Category: Privacy & Security Risk Assessment** | | |
| **Emerging Technology Privacy Risks** | I1–I4, I6, S1–S61 | Count: 47; Percentage: 6.58% | Risks related to the privacy and security implications of rapidly evolving technologies—especially AI and quantum systems—including the lack of mature standards, ethical uncertainty, and risks from novel data processing capabilities. | P |
| **Sensitive Data Exposure in Training Models** | I1, I5–I6, S1–S7, S10–S19, S21–S23, S25–S36, S38–S44, S46–S61 | Count: 27; Percentage: 3.78% | Risks of privacy breaches and security vulnerabilities due to exposure of personal or sensitive data during AI model training, including memorisation, leakage via outputs, or weak training data protections. | S |
| **Post-Quantum Cryptography Risks** | I2, I3, I5, I6, S1, S3–S5, S7, S11–S14, S20, S22–S23, S25, S28–S34, S37, S40–S41, S43, S44, S48, S51, S54, S56–S59 | Count: 23; Percentage: 3.22% | Risks associated with the security and privacy limitations of current cryptographic methods in the face of quantum computing, and uncertainties around post-quantum cryptographic algorithm adoption, implementation, and resilience. | S |
| **Emerging AI Threats & Data Vulnerabilities** | I1, I4, S1, S4, S6, S9, S10, S12–S14, S19, S21, S25, S27, S29, S33, S34, S36, S40, S43, S46–S47, S50, S52, S54, S57, S59 | Count: 23; Percentage: 3.22% | Threats from the use or misuse of AI that introduce new attack surfaces or amplify privacy risks, such as adversarial inputs, model inversion, data poisoning, and unauthorised inference. | S |
| | | **Category: Privacy & Security Controls Implementation** | | |
| **Weak Access Controls & Data Security Risks** | I1, I3, I5, I6, S1–S19, S21–S29, S30–S40, S41–S61 | Count: 36; Percentage: 5.04% | Risks stemming from inadequate security measures, such as weak authentication, insufficient authorisation, or poor encryption, leading to unauthorised access or exposure of private data. | S |
| **AI Privacy Control Implementation Gaps** | I1, I3, I5, S1–S3, S5–S7, S9–S16, S17–S19, S21–S29, S30–S40, S41–S61 | Count: 24; Percentage: 3.36% | Gaps in the design or enforcement of privacy-related controls (e.g., anonymisation, data minimisation, consent tracking) within AI workflows, which create both privacy vulnerabilities and potential security loopholes. | P |
| **Technical Privacy Control Gaps** | S2, S6, S9, S14–S19, S22–S24, S34–S40, S42–S50 | Count: 24; Percentage: 3.36% | Shortcomings in technical safeguards meant to enforce privacy—such as differential privacy, homomorphic encryption, or secure federated learning—which can also weaken system security if poorly implemented or misconfigured. | P |
| **Privacy Safeguard Deficiencies in AI** | I1, I4, I5, I6, S3, S5, S6, S9, S10, S14, S15, S17, S19, S23, S28, S29, S30, S35, S38, S39, S40, S43, S44, S47, S49, S50, S51, S54, S59 | Count: 23; Percentage: 3.22% | Risks from insufficient or poorly integrated privacy protection mechanisms in AI systems, including both organisational policy gaps and technical safeguards, potentially compromising both user privacy and system security. | P |
| | | **Category: User-Centric Privacy Considerations** | | |
| **Transparency & User Rights Risks** | I1, I2, I4, I5, I6, S1, S3–S10, S11–S13, S14–S19, S20–S22, S23–S29, S30–S36, S37–S42, S43–S61 | Count: 49; Percentage: 6.86% | Risks from insufficient transparency about how user data is handled, leading to violations of user rights. | P |
| **Bias & Fairness in AI Privacy** | I5, I6, S2, S7–S9, S12, S14–S18, S21–S22, S26–S28, S30, S33–S38, S40–S42, S43–S61 | Count: 49; Percentage: 6.86% | Risks from AI systems that introduce bias or unfairness in data handling, impacting user privacy. | P |
| **User Consent & Control Mechanisms** | I1, I2, I4, I5, I6, S2–S3, S4–S6, S7–S19, S20–S29, S30–S36, S37–S42, S43–S61 | Count: 47; Percentage: 6.58% | Risks from inadequate mechanisms for users to give informed consent and control how their data is used. | P |
| **Informed Consent Deficiencies** | S2–S4, S7–S15, S17–S22, S23–S31, S32–S39, S40–S48, S49–S61 | Count: 42; Percentage: 5.88% | Risks from inadequate processes for obtaining informed consent, potentially violating privacy rights. | P |
| **Loss of User Control Risks** | I2, I5, I6, S2, S5–S17, S19–S61 | Count: 39; Percentage: 5.46% | Risks from users losing control over their personal data when interacting with AI systems. | P |
| **User Consent & Control Risks** | I2, I3, I4, I5, I6, S2, S4, S6, S8–S11, S13–S17, S19, S22–S25, S27–S36, S37, S41, S43, S44, S47, S51, S53, S55–S61 | Count: 38; Percentage: 5.32% | Risks from lack of user control over personal data, including consent and data access management. | P |
| | | **Category: Continuous Monitoring & Improvement** | | |
| **Compliance Monitoring Gaps** | I5, S2–S10, S14–S19, S21, S23–S24, S29–S53 | Count: 23; Percentage: 3.22% | Risks due to insufficient monitoring of compliance with privacy regulations, leading to potential violations. | P |
| **Ongoing AI Auditing & Bias Mitigation** | I1, I4, I5, I6, S1–S15, S17–S19, S21–S28, S29–S61 | Count: 23; Percentage: 3.22% | Risks from lack of continuous auditing and bias mitigation in AI systems that could compromise privacy. | P |
| **Monitoring Evolving AI Privacy Risks** | I1, I2, I4, I5, I6, S1–S15, S17–S22, S23, S25–S32, S33–S61 | Count: 23; Percentage: 3.22% | Risks from inadequate monitoring of new and emerging AI privacy threats, creating vulnerabilities. | P |
| **Failure to Adapt to Emerging Risks** | S2, S3, S7, S9–S10, S14–S15, S17–S19, S21, S23–S24, S29–S35, S39–S43, S44–S47, S49–S53, S57–S60 | Count: 23 Percentage: 3.22% | Risks from the inability to detect and respond to new privacy threats from emerging AI technologies. | P |